\documentclass[12pt]{article}

\usepackage{physics}
\usepackage{tikz}

\newcommand{\adag}{a^{\dag}}
\newcommand{\bdag}{b^{\dag}}

\usepackage{amssymb}
\usepackage{amsthm}
\usepackage{comment}
\usepackage{parskip}

\usepackage{wrapfig}
\usepackage{subcaption}
\usepackage{booktabs} 
\usetikzlibrary{decorations.pathmorphing}
\usepackage{threeparttable}

\usepackage{caption}
\newcommand{\probP}{\text{I\kern-0.15em P}}
\usetikzlibrary{decorations.pathmorphing, shadows.blur}
\DeclareCaptionLabelFormat{bold}{\textbf{#1 #2}}
\usepackage[pdfstartview=XYZ,
bookmarks=true,
colorlinks=true,
linkcolor=black,
urlcolor=blue,
citecolor=black,
pdftex,
bookmarks=true,
linktocpage=true, 
hyperindex=true
]{hyperref}
\definecolor{darkgreen}{RGB}{0,170,0}
\usepackage{orcidlink}
\usepackage{doi}

\begin{document}
\begin{center}
{\large\bf Bethe ansatz solution to integrable bosonic cube networks}\\
~~\\

{\large Lachlan Bennett, Phillip S. Isaac and Jon Links}\\
~~\\

School of Mathematics and Physics, The University of Queensland, \\
St Lucia QLD 4072, Australia.
\end{center}

\begin{abstract}
We study two extended Bose-Hubbard-type Hamiltonians representing bosonic networks restricted to the
graph of a cube. For both Hamiltonians, we demonstrate that Bethe ansatz methods of solution can be
employed after applying a canonical transformation of operators. We provide the resulting Bethe
ansatz equations, and corresponding formulae for states and energies of both Hamiltonians.

\end{abstract}


\section{Introduction}
\label{Intro}

Quantum integrability provides a mathematically tractable landscape for studying models in many-body
quantum mechanics that may allow for the determination of exact solutions and physically interesting
analytic properties. Through solutions of the celebrated Yang-Baxter equation, Bethe ansatz methods
can be utilised in a variety of ways to explicitly describe and analyse such solutions.  In this
context, foundational work in \cite{TYFL2015,YTFL2017,WYTLF2018} established a class of quantum
integrable systems relating to multi-well bosonic tunneling. In particular in \cite{WYTLF2018}, the
authors demonstrate that intentionally {\em breaking} the integrability of such systems in a precise
way may still allow for certain key analytic properties to be maintained, or at least inherited,
from the associated integrable system.  Thus, it may be possible to manipulate such quantum systems
to deliberately produce certain behaviours. Such a notion is at the heart of quantum control, and
quantum engineered devices (see also, for example, \cite{YT2025} and references therein).  With this
in mind, it is clear that it is important to understand the full slew of relevant quantum integrable
systems at our disposal. \\
\\
The types of quantum integrable systems of interacting bosons that we examine in the current article
were first studied in \cite{YTFL2017}, their primary feature being a network of interacting bosons
on a complete bipartite graph. Such systems were shown to be superintegrable in \cite{BFIL2024}, in
which quantum dynamics was also studied via Bethe ansatz solutions.  Relaxing the condition of
completeness on the underlying bipartite graph, generalisations of these integrable networks of
interacting bosons were also studied in \cite{ILLW2025}, where the bipartite graph only needs
satisfy a condition of symmetry in the sense discussed in \cite{CM2015}. \\
\\
In recent years, circuits have been studied in the setting of integrable quantum spin systems
due to their amenability of producing exact expressions of physical interest
\cite{CHAL2022,MV2023,VYPM2024,GP2024,PDPZ2025}. Other recent works
\cite{BCNPdAV2022,BCV2023,BCV2023b,PBCV2023,BCCV2023,BMPV2024} involve systems of (free) fermions
restricted to various networks of graphs. By constrast, the aim of our current article is to apply
Bethe ansatz techniques to two interacting bosonic network on a cube graph.
Such extended Bose-Hubbard-type systems are relevant to models of interacting cold atoms
\cite{LMSLP2009,LPS2010,LHDLMDE2016,RBMJDG2024,CBLMZ2025}. The choice of restricting the network to a
cube is motivated by our desire to demonstrate our approach to tractable, integrable systems of the type presented in
\cite{ILLW2025}. In our setting, this involves a network on a bipartite graph that is not complete,
but is symmetric, (noting that the cube graph satisfies these conditions). 
We remark that ours is not the only work involving bosonic networks on a cube. See, for instance,
the article \cite{AMN2025} (and references therein), which studies bosonic Hamiltonians restricted to
multidimensional hypercubes, albeit without particle interaction terms.

In this article, we study two Hamiltonians of interacting bosonic networks on a cube, so-called
bosonic cube networks. These Hamiltonians are presented in Section \ref{Section 2} in equations (\ref{cubeHam
1}) and (\ref{cubeHam 2}). Our approach relies on two canonical transformations applied to the
bosonic operators. These transformations are described explicitly in Section \ref{Section 3}.
Finally, in Sections \ref{Section 4} and \ref{Section 5} we apply the Bethe ansatz solutions to the
two transformed Hamiltonians.

\section{Bosonic cube models}
\label{Section 2}

\begin{figure}[h!]
\centering
\begin{tikzpicture}[scale=1.25,
  every node/.style={
    circle,
    draw=black,            
    fill=red!40,          
    inner sep=2pt,         
    font=\footnotesize,    
    text=black             
  }]

\node (L1) at (0,0)      {1};
\node (L4) at (2,0)      {2};
\node (L2) at (1,1)  {5};
\node (L3) at (3,1)  {6};
\node (L5) at (0,-2)     {3};
\node (L8) at (2,-2)     {4};
\node (L6) at (1,-1) {7};
\node (L7) at (3,-1) {8};

\draw[line width=2pt] (L1) -- (L4) -- (L8) -- (L5) -- cycle;

\draw[line width=2pt] (L2) -- (L6);
\draw[line width=2pt] (L3) -- (L7);
\draw[line width=2pt] (L1) -- (L5);
\draw[line width=2pt] (L1) -- (L2) -- (L3) -- (L4) -- cycle;
\draw[line width=2pt] (L5) -- (L6) -- (L7) -- (L8) -- cycle;

\end{tikzpicture}
\caption{Labelling of the $Q_{3}$ cube graph. The bosonic networks (\ref{cubeHam 1}) and (\ref{cubeHam 2}) are
defined on this graph: the bosons can occupy each vertex site and can tunnel to an adjacent vertex
joined by an edge.}
\label{fig1}
\end{figure}
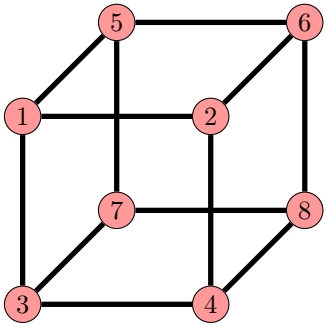

A bosonic network is a weighted graph $G=(V,E)$ equipped with a bosonic Hamiltonian $\mathcal{H}$.
Vertices $V$ label the local quantum states that bosons can occupy, and edges $E$, weighted by the
hopping amplitudes, represent the allowed transitions. The Hamiltonian $\mathcal{H}$ governs the
dynamics of the bosons within this network. A bosonic Hamiltonian is a self-adjoint operator that
depends polynomially (and non-trivially) on the creation and annihilation operators $\{ \adag_{i},
a_{i} \vert \; i=1,\ldots, \mathcal{L} < \infty\}$. The number of degrees of freedom $\mathcal{L}$
equals the number of vertices. \\
 \\
Here we introduce two non-trivial integrable bosonic networks for the cube, $Q_3= (V,E)$, with the
vertex labelling given in Fig.\ref{fig1}. This labelling is defined by the edge set
\begin{align*}
E = \big\{ (1,2),(1,3),(1,5),(2,4), (2,6), (3,4), (3,7), (4,8), (5,6), (5,7), (6,8), (7,8) \big\},
\end{align*}
where integers $1,\ldots,8$ label the vertices. Because $Q_{3}$ has eight vertices, the Hamiltonian
is expressed using the creation and annihilation operators $\{\adag_{i}, a_{i}: i=1,\ldots, 8\}$,
obeying the commutation relations $[a_{i}, \adag_{j}] = \delta_{ij}I$, $[a_{i}, a_{j}]=[\adag_{i},
\adag_{j}]=0$ for all $i,j=1,\ldots,8$. The trivial example of a cube bosonic network is the
free-boson model,
\begin{align*}
\mathcal{H}_{0} = -J\sum_{( i,j ) \in E}(\adag_{i}a_{j} + \adag_{j}a_{i}).
\end{align*}
The model lacks any interaction term between the individual particles, making it easily
diagonalisable. The difficulty is to introduce an interaction that preserves integrability. 
Two candidate Hamiltonians that address this challenge are
\begin{align}
\label{cubeHam 1}
\mathcal{H}_{1} =&\, U_0 N^{2} + U_1 N(N^{a}_{1} - N^{a}_{8} + N^{a}_{2} - N^{a}_{7}) + 4U(N^{a}_{1}
- N^{a}_{8} + N^{a}_{2} - N^{a}_{7})^2 \\
&- \dfrac{J}{2}\sum_{( i,j ) \in E}(\adag_{i}a_{j} + \adag_{j}a_{i}),\nonumber
\end{align}
and
\begin{align}
\label{cubeHam 2}
\mathcal{H}_{2} =& U_0 N^{2} + U_{1}N(N^{a}_{1} - N^{a}_{4} + N^{a}_{6}-N^{a}_{7}) + 4U(N^{a}_{1} -
N^{a}_{4} + N^{a}_{6}-N^{a}_{7})^2 \\
&- \dfrac{J}{2}\sum_{( i,j ) \in E}(\adag_{i}a_{j} + \adag_{j}a_{i}),\nonumber
\end{align}
where $N^{a}_{i} = \adag_{i}a_{i}$, $N = \sum^{8}_{i=1}N^{a}_{i}$, $U_{0}, U_1,  U \in \mathbb{R}$
set the interaction strength, and $J>0$ sets the tunnelling amplitude. Both Hamiltonians contain a
quadratic interaction term. Such quadratic interaction terms also appear in special cases of the extended
Bose-Hubbard model \cite{CBLMZ2025,HCKSM2019,CYZL2020,BHH2021,LBGCSBPLHD2022}. \\
\\

The extended Bose-Hubbard model describes key bosonic interactions and is used
in the modelling of ultra-cold quantum gases over a lattice where there are long-range dipole
interactions. For the cube, the extended Bose-Hubbard Hamiltonian is of the form 
\begin{align*}
\mathcal{H}_{eBH} = \dfrac{\widehat{U}_{0}}{2}\sum^{8}_{i=1}N^{a}_{i}(N^{a}_{i}-1) + \sum^{8}_{i\neq
j}\dfrac{\widehat{U}_{ij}}{2}N^{a}_{i}N^{a}_{j} + \sum^{8}_{i=1}\mu_{i}N^{a}_{i} -
\dfrac{J}{2}\sum_{(i,j)\in E}(\adag_{i}a_{j} + \adag_{j}a_{i}),
\end{align*}
where $\widehat{U}_{0}$ is a constant that represents the strength of the short-range interaction:
it is repulsive when $\widehat{U}_{0}>0$ and attractive when $\widehat{U}_{0}<0$.
$\widehat{U}_{ij}=\widehat{U}_{ji}$ represents the long-range interaction between sites $i$ and $j$.
Lastly, each $\mu_{i}$ is an external potential. In general, the extended Bose-Hubbard model is not
integrable. However, in the next section (Sec.\ref{Section 3}) we will show that the bosonic
networks (\ref{cubeHam 1}) and (\ref{cubeHam 2}) are integrable.

These Hamiltonians act on the Hilbert space
\begin{align}
V_{n} = \text{span}\big(\big\{ \vert n_{1},n_{2},n_{3}, n_{4}, n_{5}, n_{6},n_{7},n_{8}\rangle_{a}\;
\vert\; n_{1}+\ldots + n_{8} = n\big\}\big),
\end{align}
where the basis vectors are Fock states that represent boson occupancy at each vertex site. The Fock
states are defined as
\begin{align}
\vert n_{1},n_{2},n_{3}, n_{4}, n_{5}, n_{6},n_{7},n_{8}\rangle_{a} =
\dfrac{(\adag_{1})^{n_{1}}}{\sqrt{n_{1}!}}\dfrac{(\adag_{2})^{n_{2}}}{\sqrt{n_{2}!}}\cdots\dfrac{(\adag_{8})^{n_{8}}}{\sqrt{n_{8}!}}\vert
0 \rangle.
\end{align}
The Hilbert space dimension is 
\begin{align}
\text{dim}(V_n) = {n+7 \choose 7}.
\end{align}

\section{Canonical transformations and integrability}
\label{Section 3}

In this section, we introduce two canonical transformations on the $\adag_{i}, a_{j}$ operators in
Hamiltonians (\ref{cubeHam 1}) and (\ref{cubeHam 2}). The new basis decomposes the cube $Q_{3}$ into
simpler subgraphs that match the revised tunnelling terms. We show that the Hamiltonians can be
expressed by generators of $\mathfrak{su}(2)$. Knowing how the Hamiltonians are expressed in the
transformed basis, as well as how they are expressed in terms of $\mathfrak{su}(2)$ generators, we
find a sufficient number of symmetries to show that these models are integrable. Section
\ref{Section 4} uses these results to build the Bethe ansatz solution.

\subsection{Canonical transformation \MakeUppercase{\romannumeral 1}}

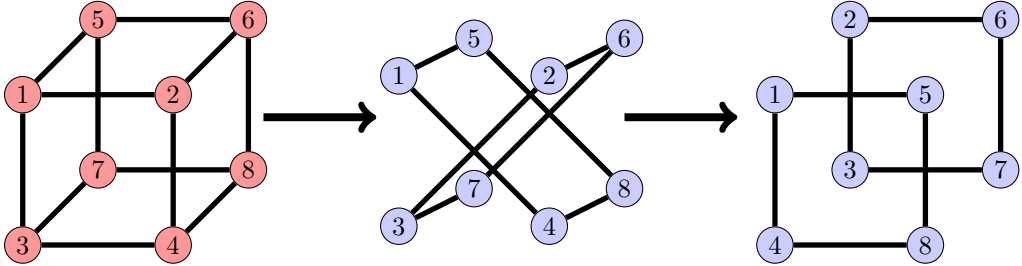
\begin{figure}[h!]
\centering
\begin{tikzpicture}[scale=1,
  every node/.style={
    circle,
    draw=black,
    fill=blue!20,
    inner sep=2pt,
    font=\footnotesize,
    text=black
  }]

\begin{scope}[every node/.append style={fill=red!40}]
\node (A1) at (0,0)   {1};   
\node (A2) at (2,0)   {2};   
\node (A3) at (0,-2)  {3};   
\node (A4) at (2,-2)  {4};   
\node (A5) at (1,1)   {5};   
\node (A6) at (3,1)   {6};   
\node (A7) at (1,-1)  {7};   
\node (A8) at (3,-1)  {8};   

\draw[line width=2pt] (A1)--(A2)--(A4)--(A3)--cycle;      
\draw[line width=2pt] (A5)--(A7) (A6)--(A8);              
\draw[line width=2pt] (A1)--(A3);                         
\draw[line width=2pt] (A1)--(A5)--(A6)--(A2);             
\draw[line width=2pt] (A3)--(A7)--(A8)--(A4);             
\end{scope}
\begin{scope}[shift={(5,0.25)}]
\node (M1) at (0,0)   {1};
\node (M2) at (2,0)   {2};
\node (M3) at (0,-2)  {3};
\node (M4) at (2,-2)  {4};
\node (M5) at (1,0.5)   {5};
\node (M6) at (3,0.5)   {6};
\node (M7) at (1,-1.5)  {7};
\node (M8) at (3,-1.5)  {8};

\draw[line width=2pt] (M1)--(M5)--(M8)--(M4)--(M1)--cycle;

\draw[line width=2pt] (M2)--(M6)--(M7)--(M3)--(M2)--cycle;
\end{scope}

\begin{scope}[shift={(10,0)}]
\node (B1) at (0,0)   {1};
\node (B2) at (2,0)   {5};
\node (B3) at (0,-2)  {4};
\node (B4) at (2,-2)  {8};
\node (B5) at (1,1)   {2};
\node (B6) at (3,1)   {6};
\node (B7) at (1,-1)  {3};
\node (B8) at (3,-1)  {7};

\draw[line width=2pt] (B1)--(B2)--(B4)--(B3)--cycle;   
\draw[line width=2pt] (B5)--(B6)--(B8)--(B7)--(B5)--cycle;           
\draw[line width=2pt] (B1)--(B3);                      
\end{scope}

\draw[->,line width=3pt] (3.2,-0.3) -- (4.7,-0.3);   
\draw[->,line width=3pt] (8,-0.3) -- (9.5,-0.3);   
\end{tikzpicture}
\caption{Transformation \MakeUppercase{\romannumeral 1} applied to the cube~$Q_{3}$.  Vertices shown
in red indicate sites in the original $a$-basis, while vertices in light blue mark the transformed
$b$-basis.  In this new basis the tunnelling network decomposes into two disjoint squares, 
matching the connectivity of Hamiltonian (\ref{new cubeHam1}). The corresponding Bethe ansatz solution is
presented in Section \ref{Section 4}.}
\label{fig2}
\end{figure}

The first canonical transformation is applied to the Hamiltonian (\ref{cubeHam 1}) and is defined:
\begin{align*}
    b_{1} &= \dfrac{1}{\sqrt{2}}(a_{1}+a_{2}), \;\; b_{2} = \dfrac{1}{\sqrt{2}}(a_{1}-a_{2}),\:\:
    b_{3} = \dfrac{1}{\sqrt{2}}(a_{3}+a_{4}), \;\; b_{4} = \dfrac{1}{\sqrt{2}}(a_{3}-a_{4}),\\
    b_{5} &= \dfrac{1}{\sqrt{2}}(a_{5}+a_{6}), \;\; b_{6} = \dfrac{1}{\sqrt{2}}(a_{5}-a_{6}),\;\;
    b_{7} = \dfrac{1}{\sqrt{2}}(a_{7}+a_{8}), \;\; b_{8} = \dfrac{1}{\sqrt{2}}(a_{7}-a_{8}).
\end{align*}
We will refer to this canonical transformation as \textit{transformation
\MakeUppercase{\romannumeral 1}}. Applying transformation \MakeUppercase{\romannumeral 1} to
Hamiltonian (\ref{cubeHam 1}) we obtain
\begin{align}
\label{new cubeHam1}
\mathcal{H}_{1} = & U_{0}N^{2} + U_1 N (N^{b}_{1}-N^{b}_{8}+N^{b}_{2} - N^{b}_{7})
+4U(N^{b}_{1}-N^{b}_{8}+N^{b}_{2} - N^{b}_{7})^{2}\\
-& J\big[(b^{\dagger}_{1}b_{4} +
b^{\dagger}_{4}b_{1})+(b^{\dagger}_{1}b_{5}+b^{\dagger}_{5}b_{1})+(b^{\dagger}_{5}b_{8}+b^{\dagger}_{8}b_{5})+(b^{\dagger}_{4}b_{8}+b^{\dagger}_{8}b_{4})+\nonumber\\&\quad(b^{\dagger}_{2}b_{6}+b^{\dagger}_{6}b_{2})+(b^{\dagger}_{3}b_{7}+b^{\dagger}_{7}b_{3})
+ (b^{\dagger}_{2}b_{3} + b^{\dagger}_{3}b_{2}) + (b^{\dagger}_{6}b_{7}+b^{\dagger}_{7}b_{6})
\big],\nonumber
\end{align}
where $N^{b}_{i} = \bdag_{i}b_{i}$. Note that $N^{a}_{1}-N^{a}_{8}+N^{a}_{2}-N^{a}_{7}$ is preserved
by transformation \MakeUppercase{\romannumeral 1}, i.e.
$$
N^{a}_{1}-N^{a}_{8}+N^{a}_{2}-N^{a}_{7} = N^{b}_{1}-N^{b}_{8}+N^{b}_{2}-N^{b}_{7}.
$$
Considering the tunnelling terms in (\ref{cubeHam 1}), we have a representation of the bosonic
network on the new basis, not as a cube but as two disjoint squares (see Fig.\ref{fig2}). We can
represent the Hilbert space as
\begin{align*}
V_{n} = \text{span}\big(\big\{ \vert n_{1},n_{2},n_{3}, n_{4}, n_{5}, n_{6},n_{7},n_{8}\rangle_{b}\;
\vert\; n_{1}+\ldots + n_{8} = n\big\}\big),
\end{align*}
where the Fock states are defined by
\begin{align}
\vert n_{1},n_{2},n_{3}, n_{4}, n_{5}, n_{6},n_{7},n_{8}\rangle_{b} =
\dfrac{(\bdag_{1})^{n_{1}}}{\sqrt{n_{1}!}}\dfrac{(\bdag_{2})^{n_{2}}}{\sqrt{n_{2}!}}\cdots\dfrac{(\bdag_{8})^{n_{8}}}{\sqrt{n_{8}!}}\vert
0 \rangle.
\end{align}
We represent the bosonic network in Fig.\ref{fig2} transformed into the new basis, where the
vertices now correspond to the new Fock state modes.

\subsection{Canonical transformation \MakeUppercase{\romannumeral 2}}

\begin{figure}[h!]
\centering
\begin{tikzpicture}[scale=1,
  every node/.style={
    circle,
    draw=black,
    fill=blue!30,
    inner sep=2pt,
    font=\footnotesize,
    text=black
  }]

\begin{scope}[every node/.append style={fill=red!40}]
\node (A1) at (0,0)   {1};   
\node (A2) at (2,0)   {2};   
\node (A3) at (0,-2)  {3};   
\node (A4) at (2,-2)  {4};   
\node (A5) at (1,1)   {5};   
\node (A6) at (3,1)   {6};   
\node (A7) at (1,-1)  {7};   
\node (A8) at (3,-1)  {8};   

\draw[line width=2pt] (A1)--(A2)--(A4)--(A3)--cycle;   
\draw[line width=2pt] (A5)--(A7);                      
\draw[line width=2pt] (A6)--(A8);                      

\draw[line width=2pt] (A1)--(A3);

\draw[line width=2pt] (A1)--(A5)--(A6)--(A2);   
\draw[line width=2pt] (A3)--(A7)--(A8)--(A4);   
\end{scope}
\begin{scope}[shift={(6,0)}]
\node (B1) at (0,0)   {1};
\node (B2) at (2,0)   {2};
\node (B3) at (0,-2)  {3};
\node (B4) at (2,-2)  {4};
\node (B5) at (1,1)   {5};
\node (B6) at (3,1)   {6};
\node (B7) at (1,-1)  {7};
\node (B8) at (3,-1)  {8};

\draw[line width=2pt] (B1)--(B2)--(B4)--(B3)--cycle;   
\draw[line width=2pt] (B5)--(B7);                      
\draw[line width=2pt] (B6)--(B8);                      
\draw[line width=2pt] (B1)--(B3);
\end{scope}

\draw[->,line width=3pt] (3.5,-0.3) -- (5.5,-0.3);

\end{tikzpicture}
\caption{Transformation \MakeUppercase{\romannumeral 2} of the cube.  Again, red vertices correspond
to the original $a$-basis; vertices in dark blue label the transformed $\widetilde{b}$-basis.
The resulting graph decomposes into one square and two dimers, in agreement with
the structure of Hamiltonian (\ref{new cubeHam2}). The corresponding Bethe ansatz solution is
presented in Section \ref{Section 5}.}
\label{fig3}
\end{figure}
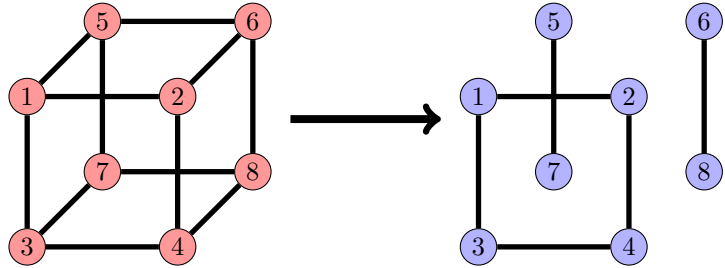

The second canonical transformation is applied to the Hamiltonian (\ref{cubeHam 2}) and is defined:
\begin{align*}
\widetilde{b}_{1} &= \dfrac{1}{\sqrt{2}}(a_{1}+a_{6}), \;\; \widetilde{b}_{6} = \dfrac{1}{\sqrt{2}}(a_{1}-a_{6}),\;\;
\widetilde{b}_{5} = \dfrac{1}{\sqrt{2}}(a_{5}+a_{2}), \;\; \widetilde{b}_{2} = \dfrac{1}{\sqrt{2}}(a_{5}-a_{2}),\\
\widetilde{b}_{3} &= \dfrac{1}{\sqrt{2}}(a_{3}+a_{8}), \;\; \widetilde{b}_{8} = \dfrac{1}{\sqrt{2}}(a_{3}-a_{8}),\;\;
\widetilde{b}_{7} = \dfrac{1}{\sqrt{2}}(a_{7}+a_{4}), \;\; \widetilde{b}_{4} = \dfrac{1}{\sqrt{2}}(a_{7}-a_{4}).
\end{align*}
We will refer to this canonical transformation as \textit{transformation
\MakeUppercase{\romannumeral 2}}. Applying transformation \MakeUppercase{\romannumeral 2} to
Hamiltonian (\ref{cubeHam 1}) we obtain
\begin{align}
\label{new cubeHam2}
\mathcal{H}_{2} = & U_{0}N^{2} + U_1 N
(\widetilde{N}^{b}_{1}-\widetilde{N}^{b}_{4}+\widetilde{N}^{b}_{6} - \widetilde{N}^{b}_{7}) +
4U(\widetilde{N}^{b}_{1}-\widetilde{N}^{b}_{4}+\widetilde{N}^{b}_{6} - \widetilde{N}^{b}_{7})^{2}\\
-& J\big[(\widetilde{b}^{\dagger}_{1}\widetilde{b}_{2} +
\widetilde{b}^{\dagger}_{2}\widetilde{b}_{1})+(\widetilde{b}^{\dagger}_{1}\widetilde{b}_{3}+\widetilde{b}^{\dagger}_{3}\widetilde{b}_{1})+(\widetilde{b}^{\dagger}_{3}\widetilde{b}_{4}+\widetilde{b}^{\dagger}_{4}\widetilde{b}_{3})+(\widetilde{b}^{\dagger}_{2}\widetilde{b}_{4}+\widetilde{b}^{\dagger}_{4}\widetilde{b}_{2})\nonumber\\
+&(\widetilde{b}^{\dagger}_{5}\widetilde{b}_{7}+\widetilde{b}^{\dagger}_{7}\widetilde{b}_{5})+(\widetilde{b}^{\dagger}_{6}\widetilde{b}_{8}+\widetilde{b}^{\dagger}_{8}\widetilde{b}_{6})
\big].\nonumber
\end{align}
For transformation \MakeUppercase{\romannumeral 2} we find that 
\begin{align*}
N^{a}_{1}-N^{a}_{4}+N^{a}_{6} - N^{a}_{7} =
\widetilde{N}^{b}_{1}-\widetilde{N}^{b}_{4}+\widetilde{N}^{b}_{6} - \widetilde{N}^{b}_{7}.
\end{align*}
The graph shown in Fig.\ref{fig3} representing the tunnelling of the bosonic network in the
transformed basis is decomposed into three disjoint graphs. Again, we have a square graph
but now with two line (i.e. dimer) graphs. We can represent the Hilbert space as
\begin{align*}
V_{n} = \text{span}\big(\big\{ \vert n_{1},n_{2},n_{3}, n_{4}, n_{5},
n_{6},n_{7},n_{8}\rangle_{\widetilde{b}}\; \vert\; n_{1}+\ldots + n_{8} = n\big\}\big),
\end{align*}
where the Fock states are defined by
\begin{align}
\vert n_{1},n_{2},n_{3}, n_{4}, n_{5}, n_{6},n_{7},n_{8}\rangle_{\widetilde{b}} =
\dfrac{(\widetilde{b}^{\dagger}_{1})^{n_{1}}}{\sqrt{n_{1}!}}\dfrac{(\widetilde{b}^{\dagger}_{2})^{n_{2}}}{\sqrt{n_{2}!}}\cdots\dfrac{(\widetilde{b}^{\dagger}_{8})^{n_{8}}}{\sqrt{n_{8}!}}\vert
0 \rangle.
\end{align}

\subsection{Symmetries}
\label{symmetries}

To show that these models are integrable, we require at least eight conserved quantities for each
model that mutually commute.

\subsubsection{The double square model}
\mbox{}\\[1.5ex]
\noindent
Let us start with the first model, we will call the \textit{double square model}. The trivial
conserved quantities are the Hamiltonian itself, $\mathcal{H}_{1}$, and the total number operator
$N$. Considering the form of the Hamiltonian (\ref{new cubeHam1}) when transformation
\MakeUppercase{\romannumeral 1} is applied, we identify a total of five conserved quantities,
\begin{align*}
\mathcal{H}_{1}, N, N^{b}_{2}+N^{b}_{6}+ N^{b}_{3}+N^{b}_{7}, N^{b}_{4} + N^{b}_{5} -
b^{\dagger}_{4}b_{5} - b^{\dagger}_{5}b_{4},N^{b}_{3}+N^{b}_{6}-b^{\dagger}_{3}b_{6} -
b^{\dagger}_{6}b_{3},
\end{align*}
which all mutually commute. To find the remaining symmetries we introduce two commuting
$\mathfrak{su}(2)$ algebras generating $\mathfrak{o}(4) \cong \mathfrak{su}(2) \oplus
\mathfrak{su}(2)$:
\begin{align*}
e_1 &= b_1^\dagger b_4 + b_1^\dagger b_5 + b_4^\dagger b_8 + b_5^\dagger b_8,\;\;
f_1 = b_4^\dagger b_1 + b_5^\dagger b_1 + b_8^\dagger b_4 + b_8^\dagger b_5,\;\;
h_1 = 2(N^{b}_1 - N^{b}_8),\\
e_2 &= b_2^\dagger b_3 + b_2^\dagger b_6 + b_3^\dagger b_7 + b_6^\dagger b_7, \;\;
f_2 = b_3^\dagger b_2 + b_6^\dagger b_2 + b_7^\dagger b_3 + b_7^\dagger b_6, \;\;
h_2 = 2(N^{b}_2 - N^{b}_7),
\end{align*}
which satisfy the commutation relations
\begin{align*}
    [e_{i},f_{j}]=\delta_{ij}h_{j}, \quad [h_{i},e_{j}]=2\delta_{ij}e_{j}, \quad [h_{i},f_{j}]=-2\delta_{ij}f_{j}.
\end{align*}
We can express the Hamiltonian (\ref{new cubeHam1}) in terms of algebra generators,
\begin{align}
    \mathcal{H}_{1} = U_{0}N^{2} + U_1 N(h_1 + h_2) + U(h_{1}+h_{2})^{2} - J(e_{1}+e_{2}+f_{1}+f_{2}).
\end{align}
This means the Casimir invariants
\begin{align*}
     C_{1} = \dfrac{1}{2}h_{1}^{2}+e_{1}f_{1}+f_{1}e_{1},\quad C_{2}=\dfrac{1}{2}h_{2}^{2} + e_{2}f_{2}+f_{2}e_{2},
\end{align*}
are also conserved quantities. The last symmetry emerges when we consider the $\mathfrak{su}(2)$ algebra generators
\begin{align*}
    E = e_{1} + e_{2} , \quad F=f_{1}+f_{2} , \quad H=h_{1}+h_{2}, 
\end{align*}
and express the Hamiltonian as
\begin{align}
\label{Spin decomp Hamiltonian 1}
    \mathcal{H}_{1} = U_{0}N^{2} + U_1 N H + 4UH^{2} -J(E+F).
\end{align}
The Casimir invariant
\begin{align*}
    C_{1+2} = \dfrac{1}{2}H^{2}+EF+FE
\end{align*}
is the eighth conserved quantity, proving integrability.

\subsubsection{The double dimer and square model}
\mbox{}\\[1.5ex]
\noindent
For the second model, the \textit{double dimer and square model}, we introduce three commuting
$\mathfrak{su}(2)$ algebras:
\begin{align*}
\widetilde{e}_1 &= \widetilde{b}_1^\dagger \widetilde{b}_2 + \widetilde{b}_1^\dagger \widetilde{b}_3
+ \widetilde{b}_2^\dagger \widetilde{b}_4 + \widetilde{b}_3^\dagger \widetilde{b}_4,\quad
\widetilde{f}_1 = \widetilde{b}_2^\dagger \widetilde{b}_1 + \widetilde{b}_3^\dagger \widetilde{b}_1
+ \widetilde{b}_4^\dagger \widetilde{b}_2 + \widetilde{b}_4^\dagger \widetilde{b}_3, \quad
\widetilde{h}_1 = 2(\widetilde{N}^{b}_1 - \widetilde{N}^{b}_4),\\
\widetilde{e}_2 &=  \widetilde{b}_5^\dagger \widetilde{b}_7 + \widetilde{b}_6^\dagger
\widetilde{b}_8, \quad \widetilde{f}_2 =  \widetilde{b}_7^\dagger \widetilde{b}_5 +
\widetilde{b}_8^\dagger \widetilde{b}_6, \quad \widetilde{h}_2 = \widetilde{N}^{b}_5 -
\widetilde{N}^{b}_7 + \widetilde{N}^{b}_6 - \widetilde{N}^{b}_8,\\
\widetilde{e}_3 &= \widetilde{b}_5^\dagger \widetilde{b}_6 + \widetilde{b}_7^\dagger
\widetilde{b}_8, \quad \widetilde{f}_3 = \widetilde{b}_6^\dagger \widetilde{b}_5 +
\widetilde{b}_6^\dagger \widetilde{b}_7, \quad \widetilde{h}_3 = \widetilde{N}^{b}_5 -
\widetilde{N}^{b}_6 + \widetilde{N}^{b}_7 - \widetilde{N}^{b}_8.
\end{align*}
We express the Hamiltonian (\ref{new cubeHam2}) in terms of algebra generators,
\begin{align}
\mathcal{H}_{2} = U_{0}N^{2} +
U_{1}N(\widetilde{h}_{1}+\widetilde{h}_{2}-\widetilde{h}_{3})+4U(\widetilde{h}_{1}+\widetilde{h}_{2}-\widetilde{h}_{3})^{2}
- J(\widetilde{e}_{1} + \widetilde{e}_{2} + \widetilde{f}_{1} + \widetilde{f}_{2}).
\end{align}
We also define another $\mathfrak{su}(2)$ algebra via the generators
\begin{align*}
\widetilde{E} = \widetilde{e}_{1} + \widetilde{e}_{2}, \quad \widetilde{F}=\widetilde{f}_{1} +
\widetilde{f}_{2}, \quad \widetilde{H} = \widetilde{h}_{1} + \widetilde{h}_{2},
\end{align*}
and can express the Hamiltonian as
\begin{align}
\label{Spin decomp Hamiltonian 2}
\mathcal{H}_{2} = U_{0}N^{2} + U_{1}N(\widetilde{H}-\widetilde{h}_{3})
+4U(\widetilde{H}-\widetilde{h}_{3})^{2} - J(\widetilde{E}+\widetilde{F}).
\end{align}
We have the Casimir invariants
\begin{align*}
\widetilde{C}_{1} = \dfrac{1}{2}\widetilde{h}_{1} + \widetilde{e}_{1}\widetilde{f}_{1} +
\widetilde{f}_{1}\widetilde{e}_{1}, \quad \widetilde{C}_{2} = \dfrac{1}{2}\widetilde{h}_{2} +
\widetilde{e}_{2}\widetilde{f}_{2}+\widetilde{f}_{2}\widetilde{e}_{2}, \quad \widetilde{C}_{1+2} =
\dfrac{1}{2}\widetilde{H} + \widetilde{E}\widetilde{F} + \widetilde{F}\widetilde{E}.
\end{align*}
We identify the eight mutually commuting conserved quantities as
\begin{align*}
\mathcal{H}_{2}, N, \widetilde{N}^{b}_{5}+\widetilde{N}^{b}_{7},
\widetilde{N}^{b}_{6}+\widetilde{N}^{b}_{8}, \widetilde{N}^{b}_{2} + \widetilde{N}^{b}_{3} -
\widetilde{b}^{\dagger}_{2}\widetilde{b}_{3} - \widetilde{b}^{\dagger}_{3}\widetilde{b}_{2},
\widetilde{C}_{1}, \widetilde{C}_{2}, \widetilde{C}_{1+2}
\end{align*}
proving integrability. Note that $\widetilde{h}_{3}$, being a linear combination of two of these
conserved quantities, is also conserved.

\section{Double square model Bethe ansatz solution}
\label{Section 4}

This section outlines the Bethe ansatz solution for the double square model defined by Hamiltonian
(\ref{cubeHam 1}). The analysis proceeds in the transformed basis. Initially, this solution requires
the construction of an alternative basis, which we will call the \textit{recursive basis}. After
establishing the recursive basis, the Bethe ansatz solution follows immediately.

\subsection{Lowest weight states}

To construct the recursive basis, we will first treat the disjoint faces of the bosonic network
separately. Let us begin with the face $(1,4,5,8)$ (see Fig.\ref{fig2}). For now, by focusing on
this face alone we will consider the algebra generators: 
\begin{align*}
e_{1}= b_1^\dagger b_4 + b_1^\dagger b_5 + b_4^\dagger b_8 + b_5^\dagger b_8,\quad f_1 =
b_4^\dagger b_1 + b_5^\dagger b_1 + b_8^\dagger b_4 + b_8^\dagger b_5, \quad h_1 = 2(N^{b}_1 -
N^{b}_8), \\
e_\alpha = b^\dagger_1 b^\dagger_8 - b^\dagger_4 b^\dagger_5, \quad f_\alpha = b_4 b_5 - b_1 b_8,
\quad h_\alpha = N^{b}_1 + N^{b}_4 +N^{b}_5 + N^{b}_8 + 2I,
\end{align*}
where we have now introduced new $\mathfrak{su}(2)$ algebra generators $e_{\alpha}, f_{\alpha},
h_{\alpha}$, satisfying commutation relations
$$
[e_{\alpha},f_{\alpha}]=h_{\alpha} ,\quad [h_{\alpha},e_{\alpha}]=2e_{\alpha},\quad
[h_{\alpha},f_{\alpha}]=-2f_{\alpha}, 
$$
and commuting with the $\mathfrak{su}(2)$ algebra generators $e_{1}, f_{1}, h_{1}$. For integers
$\ell\ge m\ge0$, define
\begin{align}
\label{Sim lowest-weights1}
|\psi^{\,\ell}_{m}\rangle
=\sum_{k=0}^{\lfloor m/2\rfloor}
C_{\ell,m,k}\,(e_\alpha)^{k}\dfrac{1}{\sqrt{(m-2k)!}}(\bdag_{4}-\bdag_{5})^{m-2k}\dfrac{(\bdag_{8})^{\ell-m}}{\sqrt{(\ell-m)!}}\vert
0\rangle
\end{align}
where
\begin{align*}
C_{\ell,m,k}=(-1)^{k}\sqrt{\dfrac{m!}{(m-2k)!}}\dfrac{(\ell-k)!}{k!\ell!},\quad
k=0,\ldots, \lfloor m/2\rfloor.
\end{align*}
These are simultaneous lowest weight states, that is, $f_{1}\vert \psi^{\,\ell}_{m}\rangle =
f_{\alpha}\vert\psi^{\,\ell}_{m}\rangle =0$.

\subsubsection{Proof that $f_1 |\psi_m^{\,\ell}\rangle = 0$}
\mbox{}\\[1.5ex]
\noindent
To show that $f_1 |\psi_m^{\,\ell}\rangle = 0$, let us write $\vert\psi^{\,\ell}_{m}\rangle$ as
\begin{align*}
|\psi^{\,\ell}_{m}\rangle
=\sum_{k=0}^{\lfloor m/2\rfloor}
C_{\ell,m,k}\,(e_\alpha)^{k}\,
|\chi_{\ell,m,k}\rangle,
\end{align*}
where
\begin{align}
\label{chi q}
|\chi_{\ell,m,k}\rangle = \dfrac{1}{\sqrt{(m-2k)!}}(\bdag_{4} -
\bdag_{5})^{m-2k}\dfrac{(\bdag_{8})^{\ell-m}}{\sqrt{(\ell-m)!}}\vert 0 \rangle,
\end{align}
which satisfies the relation $(b_4 + b_5)\,|\chi_{\ell,m,k}\rangle = 0$. We decompose $f_1$ into two
parts:
\begin{align*}
f_1 = f_{a} + f_{b}, \quad \text{where} \quad
f_{a} = (b_4^\dagger + b_5^\dagger) b_1, \quad
f_{b} = b_8^\dagger(b_4 + b_5).
\end{align*}
Each term in $|\psi_m^{\,\ell}\rangle$ has no particles in mode 1, so $f_{a}$ gives zero.
Since $(b_4 + b_5)|\chi_{\ell,m,k}\rangle = 0$, $f_{b}$ also gives zero on every term, hence $f_1
|\psi_m^{\,\ell}\rangle = 0$.

\subsubsection{Proof that $f_\alpha |\psi_m^{\,\ell}\rangle = 0$}
\mbox{}\\[1.5ex]
\noindent
Using the identity
\begin{align*}
f_\alpha\,e_\alpha^{\,k}
   =e_\alpha^{\,k}f_\alpha-k\,e_\alpha^{\,k-1}h_\alpha-k(k-1)\,e_\alpha^{\,k-1},
\end{align*}
and the relations 
$$h_\alpha|\chi_{\ell,m,k}\rangle=(\ell-2k+2)|\chi_{\ell,m,k}\rangle, \quad
f_\alpha|\chi_{\ell,m,k}\rangle=-\sqrt{(m-2k)(m-2k-1)}|\chi_{\ell,m,k+1}\rangle,$$ 
one finds
{\footnotesize
\begin{align*}
{\normalsize f_{\alpha}|\psi^{\,\ell}_{m}\rangle} =& \sum^{\lfloor m/2
\rfloor}_{k=0}C_{\ell,m,k}f_{\alpha}(e_{a})^{k}|\chi_{\ell,m,k\rangle},\\
=& -\sum^{\lfloor m/2
\rfloor}_{k=0}C_{\ell,m,k}\sqrt{(m-2k)(m-2k-1)}(e_{\alpha})^{k}|\chi_{\ell,m,k+1}\rangle \\
&- \sum^{\lfloor
m/2\rfloor}_{k=1}C_{\ell,m,k}k(\ell-k+1)(e_{\alpha})^{k-1}|\chi_{\ell,m,k}\rangle,\\
=& -\sum^{\lfloor
m/2\rfloor}_{k=1}(-1)^{k-1}\sqrt{\dfrac{m!}{(m-2k+2)!}}\dfrac{(\ell-k+1)!}{(k-1)!\ell!}\sqrt{(m-2k+2)(m-2k+1)}(e_{\alpha})^{k-1}|\chi_{\ell,m,k}\rangle\\
&- \sum^{\lfloor
m/2\rfloor}_{k=1}(-1)^{k}\sqrt{\dfrac{m!}{(m-2k)!}}\dfrac{(\ell-k)!}{k!\ell!}k(\ell-k+1)(e_{\alpha})^{k-1}|\chi_{\ell,m,k}\rangle\\
=& {\normalsize 0}.
\end{align*}}

\subsection{Recursive basis}
\label{Recursive basis 1}

Having introduced the simultaneous lowest weight states, $|\psi^{\,\ell}_{m}\rangle$, we can now define
the recursive basis. Note that we are still only considering the single face $(1,4,5,8)$. First of
all, set
\begin{align*}
|0,0,\ell,m\} = |\psi^{\,\ell}_m\rangle.    
\end{align*}
We have the actions
\begin{align*}
f_{\alpha}|0,0,\ell,m\}=0, \quad h_{\alpha}|0,0,\ell,m\}=(\ell+2)|0,0,\ell,m\}.
\end{align*}
Now, for an index $\mathcal{N} \in 2\mathbb{Z_{\text{$\geq 0$}}}$, set the recursive definition
\begin{align*}
|\mathcal{N}+2,0,\ell,m\}=
\frac{-4}
{(\mathcal{N}+2\ell+4)}e_{\alpha}|\mathcal{N},0,\ell,m\} ,
\end{align*}
and for $k=0,\dots,2(\ell-m)-1$
\begin{align*}
|\mathcal{N},k+1,\ell,m\}=\frac{1}{2(\ell-m)-k}e_{1}|\mathcal{N},k,\ell,m\}.
\end{align*}
For a state $|\mathcal{N},k,\ell,m\}$, the total number of bosons is $\mathcal{N}+\ell=n$. For the
Bethe ansatz, we need to consider the following actions:
\begin{align*}
e_{1}\,|\mathcal{N},k,\ell,m\}&=(2(\ell-m)-k)|\mathcal{N},k+1,\ell,m\},\\
f_{1}\,|\mathcal{N},k,\ell,m\}&=k|\mathcal{N},k-1,\ell,m\},\\
h_{1}\,|\mathcal{N},k,\ell,m\}&=2(k-\ell+m)|\mathcal{N},k,\ell,m\}.
\end{align*}

\subsubsection{Counting modules}
\mbox{}\\[1.5ex]
\noindent
\label{counting modules 1}
We now want to count the dimension of the Hilbert space spanned by all vectors
$|\mathcal{N},k,\ell,m\}$, where $\mathcal{N}+\ell=n$ and $k=0,\ldots, 2(\ell-m)$. We can label each
module by $\mathcal{V}(\ell-m)$, each having a dimension of $2(\ell-m)+1$. Each vector
$|\mathcal{N},0,\ell,m\}$ is a lowest weight state. We count the basis vectors with the summation:
$$ \sum_{[\mathcal{N}+\ell=n]}\sum^{l}_{m=0}[2(\ell-m)+1],$$
where we sum over all combinations of $\mathcal{N}+\ell=n$ with $\mathcal{N}$ being even. Consider
the summation
\begin{align*}
\sum^{l}_{m=0}[2(\ell-m)+1] 
= \ell(\ell+1) + \ell+1 = (\ell+1)^{2}.
\end{align*}
Let $\ell=n-\mathcal{N}$ and $\mathcal{N}=2\epsilon$, $\epsilon = 0,1,\ldots, \lfloor n/2\rfloor$, 
$$ 
\sum_{[\mathcal{N}+\ell=n]}(\ell+1)^{2} = \sum^{\lfloor n/2\rfloor}_{\epsilon
=0}(n+1-2\epsilon)^{2}.
$$
Setting $M=\lfloor n/2 \rfloor$ and using the identities
$$
\sum^{M}_{\epsilon=0}\epsilon= \dfrac{M(M+1)}{2},\quad \sum^{M}_{\epsilon=0}\epsilon^{2}=
\dfrac{M(M+1)(2M+1)}{6}.
$$
we obtain
\begin{align*}
\sum^{M}_{\epsilon=0}(n+1-2\epsilon)^{2} 
&= \sum^{M}_{\epsilon=0}\big[(n+1)^{2} - 4\epsilon(n+1) +4\epsilon^{2}\big] = {n+3 \choose 3}.
\end{align*}
This is the dimension of the Hilbert space spanned by the recursive basis for a single square face.

\subsubsection{Total Hilbert space}
\mbox{}\\[1.5ex]
\noindent
So far, we have only considered the face $(1,4,5,8)$. However, it is not difficult to construct a
basis for the whole model when we observe that a recursive basis for the face $(2,3,6,7)$ can be
constructed in the same manner with a relabeling. That is, we can consider the algebra generators
\begin{align*}
e_2 = b_2^\dagger b_3 + b_2^\dagger b_6 + b_3^\dagger b_7 + b_6^\dagger b_7, \;\;
f_2 = b_3^\dagger b_2 + b_6^\dagger b_2 + b_7^\dagger b_3 + b_7^\dagger b_6, \;\;
h_2 = 2(N^{b}_2 - N^{b}_7),\\
e_\beta = b^\dagger_2 b^\dagger_7 - b^\dagger_3 b^\dagger_6, \quad f_\beta = b_3 b_6 - b_2 b_7,
\quad h_\beta = N_2 + N_3 +N_6 + N_7 + 2I,
\end{align*}
and the simultaneous lowest weight states
\begin{align*}
\label{Sim lowest-weights2}
|\widehat{\psi}^{\,\ell}_{m}\rangle
=\sum_{k=0}^{\lfloor m/2\rfloor}
C_{\ell,m,k}\,(e_\beta)^{k}\dfrac{1}{\sqrt{(m-2k)!}}(\bdag_{3}-\bdag_{6})^{m-2k}\dfrac{(\bdag_{7})^{\ell-m}}{\sqrt{(\ell-m)!}}\vert
0\rangle.
\end{align*}
We proceed to define the recursive basis in the same manner
\begin{align*}
|0,0,\ell,m\} &= |\widehat{\psi}^{\,\ell}_{m}\rangle\\
|\mathcal{N}+2,0,\ell,m\}&=\dfrac{-4}{(\mathcal{N}+2\ell+4)}e_{\beta}|\mathcal{N},0,\ell,m\},\\
|\mathcal{N},k+1,\ell,m\}&=\frac{1}{2(\ell-m)-k}e_{2}|\mathcal{N},k,\ell,m\}.
\end{align*}
To differentiate between the faces, let us label the basis vectors corresponding to the face
$(1,4,5,8)$ as $|\mathcal{N}_{\alpha}, k_{\alpha}, \ell_{\alpha}, m_{\alpha}\}_{\alpha}$, and the
basis vectors corresponding to the face $(2,3,6,7)$ as $|\mathcal{N}_{\beta}, k_{\beta},
\ell_{\beta}, m_{\beta}\}_{\beta}$. The basis vectors for the whole Hilbert space are
\begin{align*}
|\mathcal{N}_{\alpha}, k_{\alpha}, \ell_{\alpha}, m_{\alpha}\}_{\alpha}\otimes |\mathcal{N}_{\beta},
k_{\beta}, \ell_{\beta}, m_{\beta}\}_{\beta},
\end{align*}
where $\mathcal{N}_{\alpha}+\ell_{\alpha}+\mathcal{N}_{\beta}+\ell_{\beta}=n$. Setting
$\mathcal{N}_{\alpha}+\ell_{\alpha}=n_{\alpha}$, $\mathcal{N}_{\beta}+\ell_{\beta} = n_{\beta}$, we
can compute the Hilbert space dimension to be
\begin{align*}
\sum_{[n_{\alpha}+n_{\beta}=n]}{n_{\alpha} +3 \choose 3}{n_{\beta} + 3 \choose 3} = {n+7 \choose 7},
\end{align*}
which is precisely the dimension of the Hilbert space $V_{n}$, meaning we have a complete basis.

\subsection{Spin decomposition}

In the final step before the Bethe ansatz solution, consider the Hamiltonian in the form (\ref{Spin
decomp Hamiltonian 1}), introduced in Sec.\ref{symmetries}:
\begin{align*}
\mathcal{H}_{1}=U_{0}N^{2} + U_{1}NH + U\,H^{2}-J\,(E+F),
\end{align*}
which is expressed with the $\mathfrak{su}(2)$ generators
\begin{align*}
E=e_{1}+e_{2},\quad
F=f_{1}+f_{2},\quad
H=h_{1}+h_{2}.
\end{align*}
In this section we aim to introduce states in which the actions of the $\{E,F,H\}$ generators can be
well understood. We understand how the $\{e_{1}, f_{1}, h_{1}\}$ $\mathfrak{su}(2)$ generators act
on the $\lvert\mathcal{N}_{\alpha}, k_{\alpha},\ell_{\alpha},m_{\alpha}\}_{\!\alpha}$ vectors;
corresponding to the square face $(1,4,5,8)$ of the model. Furthermore, these vectors define the
$\mathfrak{su}(2)$ modules $\mathcal{V}(s_{\alpha})$, where $s_{\alpha}=\ell-m$ is the spin.
Specifically, fixing $s_{\alpha}$ and $\mathcal{N}_{\alpha}$ we define
\begin{align*}
\mathcal V(s_{\alpha}) &= \mathrm{span}\Bigl\{
\lvert\mathcal{N}_{\alpha}, k_{\alpha},\ell_{\alpha},m_{\alpha}\}_{\!\alpha}
\;\big|\;
\ell_{\alpha}-m_{\alpha}=s_{\alpha}, \ k_\alpha = 0,\ldots, 2(\ell_{\alpha}-m_{\alpha})
\Bigr\}.
\end{align*}
Likewise, we define $\mathcal{V}(s_{\beta})$ modules for the $\{e_{2},f_{2},h_{2}\}$ generators,
corresponding to the face $(2,3,6,7)$:
\begin{align*}
\mathcal V(s_{\beta}) &= \mathrm{span}\Bigl\{
\lvert\mathcal{N}_{\beta}, k_{\beta},\ell_{\beta},m_{\beta}\}_{\!\beta}
\;\big|\;
\ell_{\beta}-m_{\beta}=s_{\beta}, \ k_\beta = 0,\ldots, 2(\ell_{\beta}-m_{\beta})
\Bigr\}.
\end{align*}
The tensor product of these modules, $\mathcal{V}(s_{\alpha})\otimes \mathcal{V}(s_{\beta})$, is
used to construct the basis for the total Hilbert space. The tensor product decomposes according to
the Clebsch-Gordan decomposition
\begin{align*}
\mathcal{V}(s_{\alpha})\otimes \mathcal{V}(s_{\beta})
=\bigoplus_{j=\lvert s_{\alpha}-s_{\beta}\rvert}^{s_{\alpha}+s_{\beta}}
\mathcal{V}(j).
\end{align*}
For every $j$, choose the coupled lowest weight vector
\begin{align}
\ket{\Omega_{j}}=\sum_{q=0}^{\Lambda}(-1)^{q}\sqrt{\binom{\Lambda}{q}}|\mathcal{N}_\alpha,q,\ell_\alpha,m_\alpha\}_{\alpha}\otimes\vert
\mathcal{N}_\beta,\Lambda-q, \ell_\beta, m_\beta\}_{\beta},
\end{align}
where $\Lambda=s_{\alpha}+s_{\beta}-j$. It satisfies $F\ket{\Omega_{j}}=0$ and
$H\ket{\Omega_{j}}=-2j\ket{\Omega_{j}}$.

\subsection{Eigenvectors and eigenvalues}
\label{Eig section 1}

Finally, we outline the Bethe ansatz solution by introducing an ansatz for the $\mathcal{H}_{1}$
eigenvectors, defined within the $\mathcal{V}(j)$ modules. For a fixed $j$, inside $\mathcal{V}(j)$
define the length--$2j$ vector
\begin{align*}
\ket{\Psi_{j}} =\prod_{r=1}^{2j}(E-u_{r}I)\ket{\Omega_{j}}.
\end{align*}
Also define
\begin{align*}
\ket{\Psi_{j;(r)}}=\prod^{2j}_{k\neq r}(E-u_{k}I)\ket{\Omega_{j}}.
\end{align*}

Map the uncoupled basis ket
$|\mathcal{N}_{\alpha}, k_{\alpha}, \ell_{\alpha}, m_{\alpha}\}_{\alpha}\otimes
|\mathcal{N}_{\beta}, k_{\beta}, \ell_{\beta}, m_{\beta}\}_{\beta}$ to the monomial
$x^{\,k_{\alpha}}y^{\,k_{\beta}}$ and set
\begin{align*}
e_{1}&\longmapsto D_{x}=2s_{\alpha}\,x-x^{2}\partial_{x},&
f_{1}&\longmapsto \partial_{x},&
h_{1}&\longmapsto 2x\partial_{x}-2s_{\alpha},\\
e_{2}&\longmapsto D_{y}=2s_{\beta}\,y-y^{2}\partial_{y},&
f_{2}&\longmapsto \partial_{y},&
h_{2}&\longmapsto 2y\partial_{y}-2s_{\beta}.
\end{align*}
This map is a Lie algebra homomorphism from
$\mathfrak{su}(2)\oplus\mathfrak{su}(2)$ to differential operators on
$\mathbb C[x,y]$.  Consequently
\begin{align*}
E\longmapsto D=D_{x}+D_{y},\quad
F\longmapsto \partial_{x}+\partial_{y},\quad
H \longmapsto 2(x\partial_{x}+y\partial_{y})-2(s_{\alpha}+s_{\beta}),
\end{align*}
still satisfy $[E,F]=H$, $[H,E]=2E$ and $[H,F]=-2F$ after the
realisation.  Inside each coupled spin-$j$ module the dynamics is governed solely by
the subalgebra
$\mathrm{span}\{E,F,H\} \subset \mathrm{span}\{e_1,f_1,h_1\}\oplus\mathrm{span}\{e_2,f_2,h_2\}$.    
We can project $E,F,H$ onto single–variable differential operators:
\begin{align*}
E\to 2j z-z^{2}\dfrac{d}{dz},\quad
F \to \dfrac{d}{dz},\quad
H \to 2z\dfrac{d}{dz}-2j.
\end{align*}
With this identification, the coupled lowest weight vector
$\ket{\Omega_{j}}$ is simply the constant polynomial $1$, and all
subsequent actions on
$|\Psi_{j}\rangle$
follow from ordinary Leibniz rules in the single variable $z$. The actions are as follows:
\begin{align*}
E\ket{\Psi_{j}}
&=-\sum^{2j}_{r=1}u_{r}\ket{\Psi_{j}}
-\sum^{2j}_{r=1}u_{r}^{2}\ket{\Psi_{j;(r)}},
\\
F\ket{\Psi_{j}}
&=2j\sum^{2j}_{r=1}\ket{\Psi_{j;(r)}},
\\
H|\Psi_{j}\rangle
&= 2j|\Psi_{j}\rangle
+2\sum_{r=1}^{2j} u_{r}|\Psi_{j;(r)}\rangle, \\
H^{2}|\Psi_{j}\rangle
&= 4j^{2}|\Psi_{j}\rangle + 4(2j-1)\sum_{r=1}^{2j} u_{r}|\Psi_{j;(r)}\rangle +
8\sum^{2j}_{r=1}\sum^{2j}_{l\neq r}
\frac{u_{r}^{\,2}}{u_{r}-u_{l}}\,
|\Psi_{j;(r)}\rangle.  
\end{align*}
The action of 
\begin{align*}
\mathcal H_{1}=U_{0}N^{2}+U_{1}N H+UH^{2}-J(E+F),
\end{align*}
on the eigenvector ansatz is
\begin{align*}
\mathcal H_{1}\ket{\Psi_{j}}
&=\Bigl(U_{0}n^{2}+2U_{1}n j+4Uj^{2}+J\sum_{r=1}^{2j}u_{r}\Bigr)\ket{\Psi_{j}}\\
&\quad+\sum_{r=1}^{2j}\Bigl[ 2n u_{r}+
4U(2j-1)u_{r}-J(2j-u_{r}^{2})
+8U\!\!\sum_{\substack{l=1\\ l\neq r}}^{2j}
\frac{u_{r}^{2}}{u_{r}-u_{l}}
\Bigr]\ket{\Psi_{j;(r)}}.
\end{align*}
Requiring every coefficient of the auxiliary states $\ket{\Psi_{j;(r)}}$ to vanish gives the Bethe
equations
\begin{align}
\label{double-square-BA-equations}
\sum_{\substack{l=1\\ l\neq r}}^{2j}
\frac{2u_{r}^{2}}{u_{r}-u_{l}}
=\frac{J}{4U}\bigl(2j-u_{r}^{2}\bigr)-\dfrac{U_1}{2U}nu_r-(2j-1)u_{r},
\qquad r=1,\dots ,2j.
\end{align}
For any solution $\{u_{1},\dots ,u_{2j}\}$, the corresponding energy is
\begin{align}
\label{double-square-energies}
   \mathcal E(j)=U_{0}n^{2}+2U_{1}n j+4Uj^{2}+J\sum_{r=1}^{2j}u_{r}.
\end{align}
These formulas hold for fixed $\ell_{\alpha},\ell_{\beta},m_{\alpha},m_{\beta}$, with
$j$ taking the values
$j=\lvert s_{\alpha}-s_{\beta}\rvert,\dots ,s_{\alpha}+s_{\beta}$.

\section{Double dimer and square model Bethe ansatz solution}
\label{Section 5}

In this section, we adapt the Bethe ansatz solution introduced in Sec.\ref{Section 4} to solve for
the double dimer and square model. The idea is very similar and the procedure is almost identical,
except that a new class of lowest weight states has to be introduced to account for the double dimers.

\subsection{Lowest weight states}

To construct the recursive basis for this model, we begin again with a single face. In dealing with
the square part of the model, $(1,2,3,4)$, we find that the same form for the lowest weight states works
for this model. For this part of the model we deal with the algebra generators:
\begin{align*}
\widetilde{e}_{1}= \widetilde{b}_1^\dagger \widetilde{b}_2 + \widetilde{b}_1^\dagger \widetilde{b}_3
+ \widetilde{b}_2^\dagger \widetilde{b}_4 + \widetilde{b}_3^\dagger \widetilde{b}_4,\quad
\widetilde{f}_1 =  \widetilde{b}_2^\dagger \widetilde{b}_1 + \widetilde{b}_3^\dagger \widetilde{b}_1
+ \widetilde{b}_4^\dagger \widetilde{b}_2 + \widetilde{b}_4^\dagger \widetilde{b}_3, \quad
\widetilde{h}_1 = 2(\widetilde{N}^{b}_1 - \widetilde{N}^{b}_4), \\
\widetilde{e}_\alpha = \widetilde{b}^\dagger_1 \widetilde{b}^\dagger_4 - \widetilde{b}^\dagger_2
\widetilde{b}^\dagger_3, \quad \widetilde{f}_\alpha = \widetilde{b}_2 \widetilde{b}_3 -
\widetilde{b}_1 \widetilde{b}_4, \quad \widetilde{h}_\alpha = \widetilde{N}^{b}_1 +
\widetilde{N}^{b}_2 +\widetilde{N}^{b}_3 + \widetilde{N}^{b}_4 + 2I,
\end{align*}
The corresponding simultaneous lowest weight states are, therefore,
\begin{align*}
|\widetilde{\psi}^{\,l}_{m}\rangle
=\sum_{k=0}^{\lfloor m/2\rfloor}
C_{l,m,k}\,(\widetilde{e}_\alpha)^{k}\dfrac{1}{\sqrt{(m-2k)!}}(\widetilde{b}^{\dagger}_{2}-\widetilde{b}^{\dagger}_{3})^{m-2k}\dfrac{(\widetilde{b}^{\dagger}_{4})^{\ell-m}}{\sqrt{(\ell-m)!}}\vert
0\rangle.
\end{align*}
However, for the double dimer part of model, we are dealing with the algebra generators
\begin{align*}
\widetilde{e}_2 =  \widetilde{b}_5^\dagger \widetilde{b}_7 + \widetilde{b}_6^\dagger
\widetilde{b}_8, \quad
\widetilde{f}_2 =  \widetilde{b}_7^\dagger \widetilde{b}_5 + \widetilde{b}_8^\dagger
\widetilde{b}_6, \quad
\widetilde{h}_2 = \widetilde{N}^{b}_5 - \widetilde{N}^{b}_7 + \widetilde{N}^{b}_6 -
\widetilde{N}^{b}_8, \\
\widetilde{e}_{\beta} = \widetilde{b}^\dagger_5 \widetilde{b}^\dagger_8 - \widetilde{b}^\dagger_6
\widetilde{b}^\dagger_7, \quad \widetilde{f}_\beta = \widetilde{b}_6 \widetilde{b}_7 -
\widetilde{b}_5 \widetilde{b}_8, \quad \widetilde{h}_\beta = \widetilde{N}^{b}_5 +
\widetilde{N}^{b}_6 +\widetilde{N}^{b}_7 + \widetilde{N}^{b}_8 + 2I.
\end{align*}
The corresponding simultaneous lowest weight states are 
\begin{align*}
|\widetilde{\phi}_{n_{7}, n_{8}}\rangle =
\dfrac{(\widetilde{b}^{\dagger}_{7})^{n_{7}}}{\sqrt{n_{7}!}}\dfrac{(\widetilde{b}^{\dagger}_{8})^{n_{8}}}{\sqrt{n_{8}!}}|0\rangle.
\end{align*}
It is straightforward to check that $\widetilde{f}_{2}|\widetilde{\phi}_{n_{7}, n_{8}}\rangle =
\widetilde{f}_{\beta}|\widetilde{\phi}_{n_{7}, n_{8}}\rangle = 0$, for all $n_{7},n_{8}\in
\mathbb{Z_{+}}$.

\subsection{Recursive basis}

Having introduced the simultaneous lowest weight vectors, $|\widetilde{\phi}_{n_{7},n_{8}}\rangle$,
we can now construct the full recursive basis for the dimer face $(5,6,7,8)$.  First, set
\begin{align*}
|0,0,n_{7},n_{8}\} = |\widetilde{\phi}_{n_{7},n_{8}}\rangle.
\end{align*}
The lowest weight conditions are as follows:
\begin{align*}
\widetilde{f}_{\beta}|0,0,n_{7},n_{8}\}&=0, \quad
\widetilde{f}_{2}|0,0,n_{7},n_{8}\}=0,\\
\widetilde{h}_{\beta}|0,0,n_{7},n_{8}\}&=(n_{7}+n_{8}+2)
|0,0,n_{7},n_{8}\},\quad
\widetilde{h}_{2}|0,0,n_{7},n_{8}\}=-(n_{7}+n_{8})|0,0,n_{7},n_{8}\}.
\end{align*}
For an even index $\widetilde{\mathcal{N}}\in 2\mathbb{Z}_{\ge0}$ define
\begin{align*}
|\widetilde{\mathcal{N}}+2,0,n_{7},n_{8}\} =
\frac{-4}{\widetilde{\mathcal{N}}+2(n_{7}+n_{8})+4}
\widetilde{e}_{\beta}|\widetilde{\mathcal{N}},0,n_{7},n_{8}\},
\end{align*}
and for $k=0,\dots,n_{7}+n_{8}-1$ set
\begin{align*}
|\widetilde{\mathcal{N}},k+1,n_{7},n_{8}\} =
\frac{1}{n_{7}+n_{8}+\widetilde{\mathcal{N}}-k}
\widetilde{e}_{2}|\widetilde{\mathcal{N}},k,n_{7},n_{8}\}.
\end{align*}
For a state $|\widetilde{\mathcal{N}},k,n_{7},n_{8}\}$ (total boson number
$n=\widetilde{\mathcal{N}}+n_{7}+n_{8}$) the relevant actions are
\begin{align*}
\widetilde{e}_{2}|\widetilde{\mathcal{N}},k,n_{7},n_{8}\}
&= (n_{7}+n_{8}+\widetilde{\mathcal{N}}-k)
|\widetilde{\mathcal{N}},k+1,n_{7},n_{8}\},\\
\widetilde{f}_{2}|\widetilde{\mathcal{N}},k,n_{7},n_{8}\}
&= k|\widetilde{\mathcal{N}},k-1,n_{7},n_{8}\},\\
\widetilde{h}_{2}|\widetilde{\mathcal{N}},k,n_{7},n_{8}\}
&= (2k-n_{7}-n_{8})|\widetilde{\mathcal{N}},k,n_{7},n_{8}\},\\
\widetilde{h}_{3}|\widetilde{\mathcal{N}},k,n_7,n_8\} &= (n_7 - n_8)|\widetilde{\mathcal{N}}, k,
n_7, n_8\}.
\end{align*}

\subsubsection{Counting modules and the total Hilbert space}
\mbox{}\\[1.5ex]
\noindent
Let us count the dimension of the Hilbert space spanned by all states
$|\widetilde{\mathcal{N}},k,n_7, n_8\}$, where $\widetilde{\mathcal{N}} + n_7 + n_8 = n$ and $k=
0,\ldots, n_7 +n_8 +1$. We label each module by $\mathcal{V}((n_7+ n_8)/2)$, each having a dimension
of $n_7+n_8 +1$. Counting the basis vectors is equivalent to evaluating the sum
\begin{align*}
\sum_{\widetilde{[\mathcal{N}}+n_7+n_8=n]}(n_{7}+n_8+1),
\end{align*}
which adds the quantity $(n_7+n_8+1)$ for each combination of $\widetilde{\mathcal{N}}+n_{7}+n_8=n$.
Since $\mathcal N=n-n_{7}-n_{8}$ must be even, set $\rho=n_{7}+n_{8}=n-2\epsilon$
with $\epsilon=0,1,\ldots,\lfloor n/2\rfloor$.
For each $\rho$ there are $\rho+1$ ordered pairs
$(n_{7},n_{8})$, and for every pair the module
contributes $\rho+1$ basis vectors.  Therefore, the dimension of the Hilbert space is
\begin{align*}
\sum_{\epsilon =0}^{\lfloor n/2\rfloor}(n-2\epsilon+1)^{2}
=
\binom{\,n+3\,}{3}.
\end{align*}
Hence, the recursive basis for a single square spans the complete $n$ boson Hilbert space in four
modes, whose dimensions are the familiar tetrahedral numbers.\\
\\
We can now construct the basis vectors
\begin{align*}
|\widetilde{\mathcal{N}}_{\alpha},k_{\alpha},l,m\}_{\alpha}\otimes|\widetilde{\mathcal{N}}_{\beta},
k_{\beta}, n_7,n_8\}_{\beta},
\end{align*}
where $|\widetilde{\mathcal{N}}_{\alpha},k_{\alpha},l,m\}_{\alpha}$ are constructed from the lowest
weight states $|\widetilde{\psi}^{\,l}_{m}\rangle$ in the same manner as presented in Sec.\ref{Recursive
basis 1} and $|\widetilde{\mathcal{N}}_{\beta}, k_{\beta}, n_7,n_8\}_{\beta}$ is constructed as
shown above. These basis vectors span the total Hilbert space, which can be demonstrated using the
same argument introduced in Sec.\ref{counting modules 1}.

\subsection{Spin decomposition} 

Recall the $\mathfrak{su}(2)$ generators that appear in the
Hamiltonian (\ref{Spin decomp Hamiltonian 2}):
\begin{align*}
  \widetilde E &= \widetilde e_{1}+\widetilde e_{2}, &
  \widetilde F &= \widetilde f_{1}+\widetilde f_{2}, &
  \widetilde H &= \widetilde h_{1}+\widetilde h_{2},
\end{align*}
For the square face $(1,2,3,4)$ we construct $\mathcal{V}(s_{\alpha})$ modules accociated with a set
of states $\lvert\widetilde{\mathcal{N}}_{\alpha},\widetilde k_{\alpha},\ell,m\}_{\!\alpha}$ and a
spin $s_{\alpha}= \ell-m$. These modules have dimension $2s_{\alpha}+1$. For the dimer face
$(5,6,7,8)$ we construct $\mathcal{V}(s_{\beta})$ modules of spin $s_{\beta}= (n_{7}+n_{8})/2$ and
dimension $2s_{\beta}+1$, accociated with the $\lvert\widetilde{\mathcal{N}}_{\beta},\widetilde
k_{\beta},n_{7},n_{8}\}_{\!\beta}$ vectors. Fix $\bigl(s_{\alpha},s_{\beta}\bigr)$,
$\bigl(\widetilde{\mathcal{N}}_{\alpha}, \widetilde{\mathcal{N}}_{\beta}\bigr)$ and define
\begin{align*}
\mathcal V(s_{\alpha}) &= \mathrm{span}\Bigl\{
\lvert\widetilde{\mathcal{N}}_{\alpha},\widetilde k_{\alpha},\ell,m\}_{\!\alpha}
\;\big|\;
\ell-m=s_{\alpha}, \ k_\alpha = 0,\ldots, 2(\ell-m)
\Bigr\},\\
\mathcal V(s_{\beta}) &= \mathrm{span}\Bigl\{
\lvert\widetilde{\mathcal{N}}_{\beta},\widetilde k_{\beta},n_{7},n_{8}\}_{\!\beta}
\;\big|\;
(n_{7}+n_{8})/2=s_{\beta}, \ k_\beta = 0,\ldots, n_7+n_8
\Bigr\}.
\end{align*}
Under $\{\widetilde E,\widetilde F,\widetilde H\}$ one has the standard
Clebsch–Gordan decomposition
\begin{align*}
\mathcal V(s_{\alpha})\otimes\mathcal V(s_{\beta})
= \bigoplus_{j=\lvert s_{\alpha}-s_{\beta}\rvert}^{s_{\alpha}+s_{\beta}}
\mathcal V(j),
\end{align*}
where $\mathcal V(j)$ is an irreducible module of spin $j$. For every $j$ choose
\begin{align*}
\lvert\widetilde\Omega_{j}\rangle
= \sum_{q=0}^{\Lambda}(-1)^{q}
\sqrt{\binom{\Lambda}{q}}\;
\lvert\widetilde{\mathcal{N}}_{\alpha},q,\ell,m\}_{\!\alpha}
\otimes
\lvert\widetilde{\mathcal{N}}_{\beta},\Lambda-q,n_{7},n_{8}\}_{\!\beta},
\end{align*}
with $\Lambda=s_{\alpha}+s_{\beta}-j$; this satisfies $\widetilde
F\,\lvert\widetilde\Omega_{j}\rangle=0$ and $\widetilde
H\,\lvert\widetilde\Omega_{j}\rangle=-2j\,\lvert\widetilde\Omega_{j}\rangle$. Acting with
$\widetilde E$ repeatedly generates a complete basis for $\mathcal V(j)$.\\
\\
The commuting third Cartan element $\widetilde
h_{3}=\widetilde{N}^{b}_{5}-\widetilde{N}^{b}_{6}+\widetilde{N}^{b}_{7}-\widetilde{N}^{b}_{8}$ acts
diagonally with eigenvalue
\begin{align*}
\Delta = n_{7}-n_{8},
\end{align*}
so every state in $\mathcal V(j)$ is also an eigenvector of $\widetilde h_{3}$ with eigenvalue
$\Delta$.

\subsection{Eigenvectors and eigenvalues} \label{double dimer eigs}

Inside a fixed spin sector $\mathcal V(j)$ introduce the ansatz
\begin{align*}
|\widetilde\Psi_{j}\rangle = \prod_{r=1}^{2j}\bigl(\widetilde
E-u_{r}\,I\bigr)|\widetilde\Omega_{j}\rangle.
\end{align*}
Also define
\begin{align*}
|\widetilde\Psi_{j;(r)}\rangle =\prod_{k\neq r}(\widetilde E-u_{k})|\widetilde\Omega_{j}\rangle.
\end{align*}

Like in Sec.\ref{Eig section 1}, we realise the total generators as restricted to a single-variable
operator
\begin{align*}
\widetilde E \to 2j\,z-z^{2}\dfrac{d}{dz},\quad
\widetilde F \to \dfrac{d}{dz},\quad
\widetilde H \to 2z\dfrac{d}{dz}-2j,
\end{align*}
where $s_{\alpha} = \ell - m$ and $s_{\beta} = (n_7 + n_8)/2$.  
Using this differential realisation, we obtain the actions for
$\widetilde E$, $\widetilde F$, $\widetilde H$ on the vectors
$\ket{\Psi_{j}}$:
\begin{align*}
\widetilde E|\widetilde\Psi_{j}\rangle
&= -\sum_{r=1}^{2j}u_{r}|\widetilde\Psi_{j}\rangle
-\sum_{r=1}^{2j}u_{r}^{2}|\widetilde\Psi_{j;(r)}\rangle,\\
\widetilde F|\widetilde\Psi_{j}\rangle
&= 2j\sum_{r=1}^{2j}|\widetilde\Psi_{j;(r)}\rangle,\\
(\widetilde H- \widetilde{h}_{3})|\widetilde\Psi_{j}\rangle
&= (2j-\Delta)|\widetilde\Psi_{j}\rangle
+2\sum_{r=1}^{2j}u_{r}|\widetilde\Psi_{j;(r)}\rangle,\\
(\widetilde H-\widetilde h_{3})^{2}|\widetilde\Psi_{j}\rangle
&= (2j-\Delta)^{2}|\widetilde\Psi_{j}\rangle
+4(2j-\Delta-1)\sum_{r=1}^{2j}u_{r}|\widetilde\Psi_{j;(r)}\rangle
+8\sum^{2j}_{r=1}\sum^{2j}_{l\neq r}\frac{u_{r}^{2}}{u_{r}-u_{l}}
|\widetilde\Psi_{j;(r)}\rangle .
\end{align*}
Acting with
\begin{align*}
\mathcal{H}_{2}=U_{0}N^{2}+U_{1}N(\widetilde{H}-\widetilde{h}_{3})+4U\bigl(\widetilde H-\widetilde
h_{3}\bigr)^{2}-J(\widetilde E+\widetilde F),
\end{align*}
and collecting the independent vectors
$\{|\widetilde\Psi_{j}\rangle,|\widetilde\Psi_{j;(r)}\rangle\}$ yields
\begin{align*}
\mathcal{H}_{2}\,|\widetilde\Psi_{j}\rangle
&=\Bigl(U_{0}n^{2}+U_{1}n(2j-\Delta)+4U(2j-\Delta)^{2}
+J\sum_{r=1}^{2j}u_{r}\Bigr)|\widetilde\Psi_{j}\rangle\\
&\quad+\sum_{r=1}^{2j}\Bigl[2U_{1}n u_r + 4U(2j-\Delta-1)u_{r} -J(2j-u_{r}^{2}) +8U\sum^{2j}_{l\neq
r}\frac{u_{r}^{2}}{u_{r}-u_{l}} \Bigr]|\widetilde\Psi_{j;(r)}\rangle .
\end{align*}
Requiring the coefficients of every $|\widetilde\Psi_{j;(r)}\rangle$ to vanish gives the Bethe
ansatz equations
\begin{align}
\label{double dimer BA equations}
\sum_{\substack{l=1\\ l\neq r}}^{2j}\frac{2u_{r}^{2}}{u_{r}-u_{l}} =
\frac{J}{4U}\bigl(2j-u_{r}^{2}\bigr) - \dfrac{U_1}{2U}nu_r-(2j-\Delta-1)u_{r}, \qquad r=1,\dots ,2j.
\end{align}
Solutions $\{u_{1},\dots ,u_{2j}\}$ of these equations give eigenvectors
$|\widetilde\Psi_{j}\rangle$ with energies
\begin{align}
\label{double dimer energies}
\mathcal{E}(j,\Delta)= U_{0}n^{2} + U_{1}n(2j-\Delta) +4U(2j-\Delta)^{2}  +J\sum_{r=1}^{2j}u_{r}.
\end{align}
Here $n=\widetilde N_{\alpha}+l-m+\widetilde N_{\beta}+n_{7}+n_{8}$ is the total boson number, $j$
ranges over $\lvert s_{\alpha}-s_{\beta}\rvert,\dots ,s_{\alpha}+s_{\beta}$, and
$\Delta=n_{7}-n_{8}$ is the conserved quantum number for the dimer face. These formulas provide the
complete Bethe ansatz solution of the double dimer and square model.

\section{Conclusion}

In this article we have presented the Bethe ansatz solution of two extended Bose-Hubbard-type model
Hamiltonians representing bosonic networks on a cube graph. We solved both models using Bethe
ansatz methods after applying a canonical transformation in each case to reduce the models to a
double-square and a double-dimer and square model respectively. We remark that the family of
integrable bosonic networks presented in \cite{ILLW2025}, also includes an example on a cube. In
that case, under a transformation that diagonalises the adjacency matrix of the graph, the model
reduces to a union of disjoint dimers. The Bethe ansatz for the disjoint dimer case has been explored
recently in the article \cite{Links2025}. 

\section*{Acknowledgements}

This work was supported by the Australian Research
Council through Discovery Project DP200101339. The authors acknowledge the traditional
owners of the land on which The University of Queensland (St. Lucia campus) operates, the Turrbal and
Jagera people.

\medskip


\end{document}